\documentclass[12pt]{article}

\usepackage{geometry}
\usepackage[numbers]{natbib}
\usepackage{graphicx}
\usepackage{amsmath}
\usepackage{amsthm}

  \theoremstyle{plain}
  \newtheorem{prop}{\protect\propositionname}
  \theoremstyle{definition}
  \newtheorem{defn}{\protect\definitionname}
 \theoremstyle{definition}
  \newtheorem{example}{\protect\examplename}
\theoremstyle{plain}
\newtheorem{thm}{\protect\theoremname}
  \theoremstyle{plain}
  \newtheorem{conjecture}{\protect\conjecturename}

\usepackage{algorithm,algpseudocode}

\makeatother

\usepackage[english]{babel}
  \providecommand{\conjecturename}{Conjecture}
  \providecommand{\definitionname}{Definition}
  \providecommand{\examplename}{Example}
  \providecommand{\propositionname}{Proposition}
\providecommand{\theoremname}{Theorem}

\begin{document}
${}$\\
\uppercase{\textbf{Study of a dynamic cooperative trading queue routing control scheme for freeways and facilities with parallel queues}}\\
\\
\\
\\
\\
\\
\\
\\
\\
\\
\\
\\
\\
\textbf{Roger Lloret-Batlle}\\
PhD Student\\
University of California, Irvine\\
Institute of Transportation Studies\\
4000 Anteater Instruction and Research Bldg (AIRB)\\
Irvine, CA 92697-3600\\
Email: rlloretb@uci.edu\\
\\
\\
\textbf{R. Jayakrishnan}\\
Professor\\
University of California, Irvine\\
Institute of Transportation Studies\\
4000 Anteater Instruction and Research Bldg (AIRB)\\
Irvine, CA 92697-3600\\
Email: rjayakri@uci.edu\\
\\
\\
\\
%Word count: 6,423 words text + 4 tables/figures x 250 words (each) = 7,423 words\\
\\
\\
\\
\\
\\
\\
August 1st, 2017
\newpage

\section{Abstract}
This article explores the coalitional stability of a new cooperative
control policy for freeways and parallel queuing facilities with multiple servers.
Based on predicted future delays per queue or lane, a VOT-heterogeneous
population of agents can agree to switch lanes or queues and transfer
payments to each other in order to minimize the total cost of the
incoming platoon. The strategic interaction is captured by an n-level
Stackelberg model with coalitions, while the cooperative structure
is formulated as a partition function game (PFG). The stability concept
explored is the strong-core for PFGs which we found appropiate given
the nature of the problem. This concept ensures that the efficient
allocation is individually rational and coalitionally stable. We analyze
this control mechanism for two settings: a static vertical queue and
a dynamic horizontal queue. For the former, we first characterize
the properties of the underlying cooperative game. Our simulation
results suggest that the setting is always strong-core stable. For
the latter, we propose a new relaxation program for the strong-core
concept. Our simulation results on a freeway bottleneck with constant
outflow using Newell's car-following model show the imputations to
be generally strong-core stable and the coalitional instabilities
to remain small with regard to users' costs.\\
\\
\textit{Keywords}: lane-changing, stability, core, dynamic queue routing,
parallel queues, connected vehicles

\newpage

\section{Introduction}

Connected vehicle environments bring new opportunities in the operation
of traffic infrastructures. Until now, vehicles traveling on a link
selfishly self-organized themselves into a service order which corresponds
on average to a First-Come First-Served (FCFS) order, without any
mutual exchange of urgency information. For instance, lane changes
naturally violate the FCFS service order through a gap acceptance
distributed mechanism, with very limited or no precise future delay
information. New connectivity can increase the efficiency of lange
changes by providing better future delay estimation and by incorporating
a new variable: travellers' Value Of Time (VOT).

If it was possible for users to be informed of the downstream traffic
conditions for each lane, i.e. the downstream delay on each lane at
a particular timestep, and if they could communicate their VOT values
to each other, travelers could decide which lane changes are the most
beneficial for everybody. Thus, agents can violate the initial order
by creating coalitions which give incentives to agents in front to
choose longer queues in exchange of a side payment. This would lead
to a different level of service for each lane or queue,the less congested
lanes or queues then becoming faster than the more congested ones
and preferable for the higher-VOT travellers who may be willing to
pay. The concept presented in this article is applicable to both facilities
with parallel queues (multiple parallel servers) i.e. access gates
at ports, traffic intersections, and bottlenecks in a freeway section.
We will use the terms ``queue'' and ``lane'' indistinctively,
as the essential level of performance of each lane is captured by
the queue associated with it, for the conceptual purposes of this
paper.

This article presents a new dynamic queue routing control scheme which
violates FCFS and outperforms it in efficiency while being core-stable.
The policy is now outlined. Agents can choose which queue or lane
they want to switch to. Naturally, vehicles in front get to choose
earlier. Agents are assumed to be having perfect knowledge of the
delay per lane (or queue) as well as the delay increases due to nearby
vehicles' lane changes. Agents can communicate their values of time
to any other vehicle they want to interact with. Agents can form coalitions
and exchange payments among them to improve their utility. Our mechanism
implements the most efficient allocation and ensures that all agents
and any subset of agents present in the system are better off by participating
and cooperating with the outcome solution. This is ensured by employing
the concept of the core, the pillar of cooperative game theory. Broadly, the core is a feasible set defined by constraints which define the stability of coalitions based on their worth. We hasten to add here that such exchanges may not legally allowed in
traffic systems; however, we assume that demonstrated social/system
efficiency can lead to regulatory changes in future.

Contrary to most common applications in cooperative game theory, traffic
operations present externalities. This means that the worth of a particular
coalition depends on what other coalitions do. This brings us to the
domain of partition function games (PFG), a superset of the more commonly
used characteristic function games (CFG), which are not complex enough
to express externalities. Equivalent stability concepts to the core
are defined for partition function games. In particular, we are going
to use the strong core \citep{ChanderPFG2014}, which we believe is small
enough to be meaningful and apparently non-empty for the simplified
version of the current application. A fundamental result in \citep{ichiishi1981}
establishes an equivalence between strategic games and partition games.
This paper will actually relate both approaches, since the strategic-cooperative
interaction is modeled as the optimization of the union of n-level
Stackelberg games with coalitions.

We claim the following contributions in this article. First, we present
this novel operational scheme for parallel queues and freeway management.
Second, we are the first to use and solve a multiple-discrete-strategies
n-level Stackelberg game with coalitions. Third, we found that the
problem appears to be always strong-core stable for the vertical queue
case, and that it is generally stable for the horizontal queue case
as well. Finally, we propose a new relaxation for the strong-core
concept found in \citep{ChanderPFG2014} and generalize it to the dynamic
domain.

The article is organized as follows: section 2 presents the meaningful
literature from microeconomics on queue games and the utilized strategic
structures, section 3 presents a static version on the cooperative
queuing problem, modeling the queue routing as a parallel static
vertical queue, section 4 presents a dynamic version of the problem,
modeling the queue routing as parallel horizontal queues, and section
5 presents the conclusions and further research.

\section{Literature review}

Microeconomics literature has extensively explored the stability,
fairness and truthfulness of priority queues \citep{Chun2016book}
for single and parallel queues \citep{Chun2008}. In priority queues,
an unordered set of agents with heterogeneous values of time occupies
positions on a line valued with linear delay. The efficient queue
ordering is the one which places the agents sorted by decreasing value
of time. However, queues in transportation systems involve agents
with physical dimension and not all queue orderings are possible due
to agent obstruction.

\citep{bradford1996} studied pricing and incentives for a multiserver
queuing facility. He analyzes both social welfare versus operator's
revenue but does not enforce budget balancedness nor cooperation of
any kind. His results are based on steady state and reach standard
marginal pricing conclusions in efficiency maximization. 

\citep{Yechiali1971} examines stationary equilibria in a GI/M/1 queue policy in which users agree over a common service toll, which is later redistributed among participants. However, it does not examine coalitional stability between agents nor applies a deterministic analysis on individual vehicles as the present article does. Other authors have tackled similar queue problem on the demand side \citep{Brink2015}.

The cooperative interaction between agents for this problem is represented
by partition function games. PFG are normally classified by the externality,
either positive or negative, that a two coalitions forming creates
on a third one. \citep{Hafalir2007} explores the role of convexity
on efficiency and core stability. \citep{Abe2016} explores PFG with
either positive of negative externalities. Similarly to characteristic
function games, although less studied, several stability concepts
have been developed for this more expressive concept \citep{hartkurz1983}.
These concepts being generally too large when non-empty \citep{Chander1995,Chander1997}
propose the gamma-core and the strong-core \citep{ChanderPFG2014}.

The strategic interaction between travelers has an inherent arrival
ordering. The most adequate equilibrium concept is that of (multilevel)
Stackelberg equilibrium. Much has been said about two and three level
variants of this concept. Little has been studied on the more general
multilevel case. \citep{Bialas1989a} explore coalition formation in
level Stackelberg games for linear resource problems.

On a related application, \citep{lloretbatlle2017ISTTT} proposes
a queue jumping mechanism for general purpose freeway operations,
in which vehicles coming from upstream can pay queued vehicles for
being overtaken. Stability in the problem emanates from envy-freeness,
naturally found in position enviornments \citep{Varian2007}.

\section{Parallel vertical queues: Static problem}

We have a section of road which has a bottleneck downstream of constant
outflow, this bottleneck can either be highway congestion, a multiserver
queue from a port terminal or border crossing point or a saturated
intersection. The section downstream has $m$ lanes, and the section
upstream has $l$ lanes. $m$ may be larger, equal or smaller than
$l$. From each lane $l\in L$, a subset $M_{l}\subseteq M$ is accessible.
There is a set $N$ of vehicles approaching the queue from the back.
Each lane $l\in L$ has $N_{l}\subseteq N$ vehicles. Each downstream
lane $m$ has a queue $Q_{m}\geq0$ built up. Without loss of generality,
we assume that $Q_{m^{\prime}}\leq Q_{m},\; \forall m^{\prime}>m$. These
queues can represent actual queues of stopped vehicles or congested
traffic in the link transmission model fashion. For the analysis in
this section, the facility dispatches one vehicle per unit of time
per queue.

We decompose the ordered set of agents $N\backslash\{i\}$ in two
sets. Let $A(i)=\{k\in N\;|\;k<i\}$ be the set of predecessors of
agent $i$ and $F(i)=\{k\in N\;|\;k>i\}$ the set of followers of
$i$. Let $A(m,i)$ be the set of predecessors of $i$ which choose
lane $m$. Let $j_{m}=|A(m,i)|,\;\forall m\in M$. This defines a lane
choice set $\sigma:\;N\rightarrow M^{N}$. Contrary to priority queues,
in our model, a traveler cannot advance a predecessor unless he joins
a queue which is shorter than the queue that predecessor has joined.

The delay for agent $i\in N$ joining lane $m$ given lane choice
set $\sigma(N)$ is $d_{i}(Q,\sigma(A(i)))=(Q_{m}+j_{m}-1)$ and the
valuation experienced by agent $i$, $v_{i}(Q,\sigma(A(i)))=-\theta_{i}d_{i}(Q,\sigma(A(i)))$,
where $\theta_{i}$ represents the value of delay in monetary units
per unit of time. This variable will also be called the type of agent
$i$. Agent $i$ is charged a price $p_{i}$ for bearing the delay
$d_{i}(Q,\sigma(A(i)))$. Finally, his utility is $u_{i}=v_{i}(Q,\sigma(A(i)))-p_{i}$.

\begin{figure}
\begin{centering}
\includegraphics[width=2in]{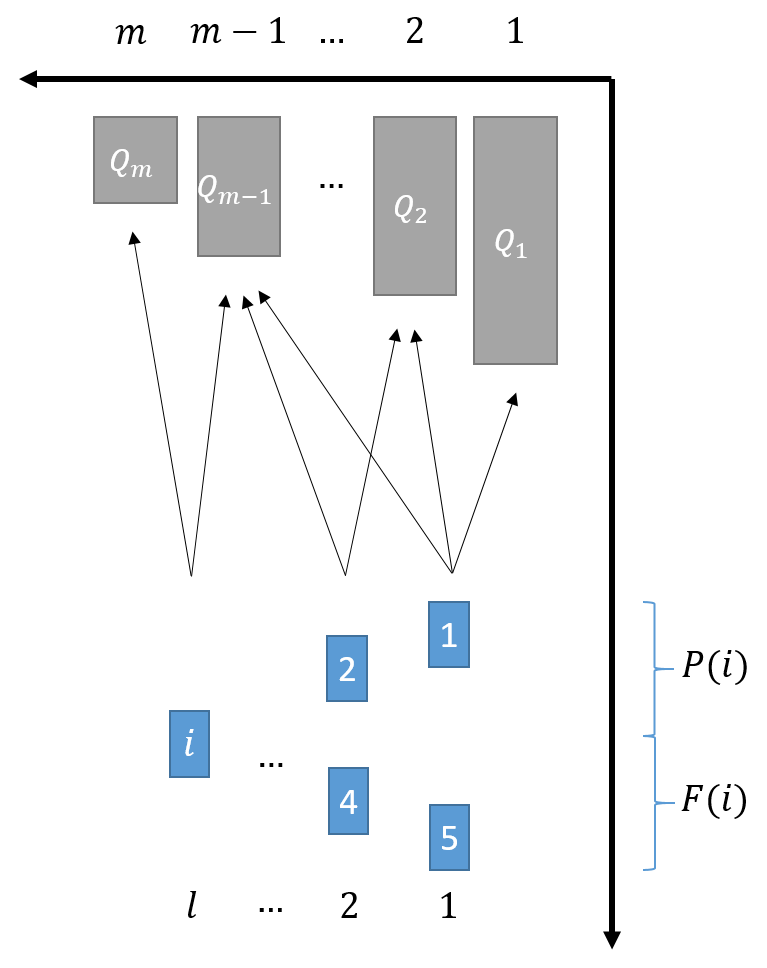}
\par\end{centering}
\caption{General problem configuration}
\end{figure}

Vehicles upstream belong to different lanes and are ordered based
on their proximity to the downstream boundary. Again without loss
of generality, we disregard the lanes $l$ where vehicles are located,
which is equivalent to assuming that $M_{l}=M$. If there is no communication
between agents, vehicles will join the downstream bottleneck on a
First-Come First-Served (FCFS) basis, each vehicle selecting the shortest
queue. It is easy to see that if the initial arrival order is not
monotonically decreasing on types, the resulting queue ordering will
be inefficient, if we view efficiency in a utilitarian social welfare
sense. However, if vehicles were to cooperate with each other, that
is, forming coalitions to alter this initial ordering, a more efficient
ordering for everyone would be achieved. This cooperation would be
on terms of multilateral agreements on which lane, every vehicle of
the coalition would choose. This defines a multilevel Stackelberg
game with coalitions. Of course, such coalition-forming would require
communications and decision-making of the kind human drivers in traffic
may not accomplish, but apps representing them can accomplish, and
the technology for it certainly exists already.

\citep{Bialas1982,Bialas1989a} explored some sufficient properties
for stability for this kind of games, but on a linear resource allocation
environment. We apply a similar recursion to our problem, but this
time for pure strategies in extensive game. This recursion will give
us the value generated by each coalition, given a particular coalition
structure (partition of $N$). The recursion starts from the last
vehicle and goes up. At each level, the user $i$ attempts to optimize
the sum of the valuation of users who are behind it, $j\in N\;|\;j>i$
and which belong to the particular coalition $i$ belongs to. The
back recursion to solve the n-level Stackelberg problem with coalitions
is, given a partition $P=\{S_{1},...,S_{j}\}\;|\;\cap_{j}S_{j}=\emptyset$:

\begin{equation}
V_{S(i)|P}^{*}(h)=\max_{k\in K_{i}(h)}\left\{ v_{i}(h,k)+V_{S(i)|P}^{*}(h\cup k)\right\} \;\forall i=n-1,...,1,\;h\in H_{i}\label{eq:recursion}
\end{equation}

\begin{equation}
V_{S(n)|P}(h)=\max_{k\in K_{n}(h)}\left\{ v_{n}(h,k)\right\} 
\end{equation}

Where $S(i)$ is the coalition where $i$ belongs.

(\ref{eq:recursion}) shows that for every user $i$, for every past
history $h\in H_{i}$ up to user $i$, the agent selects the action
$k$ belonging to the set of available actions given $h$, $K_{i}(h)$,
which maximizes the sum of two terms. The optimal value $V_{S(i)}^{*}(h\cup k)$
which emanates from the previous step $i+1$ and, $v_{i}(h,k)$, the
valuation of user $i$ from choosing action $k$, given history $h$.

\begin{figure}
\begin{centering}
\includegraphics[width=3in]{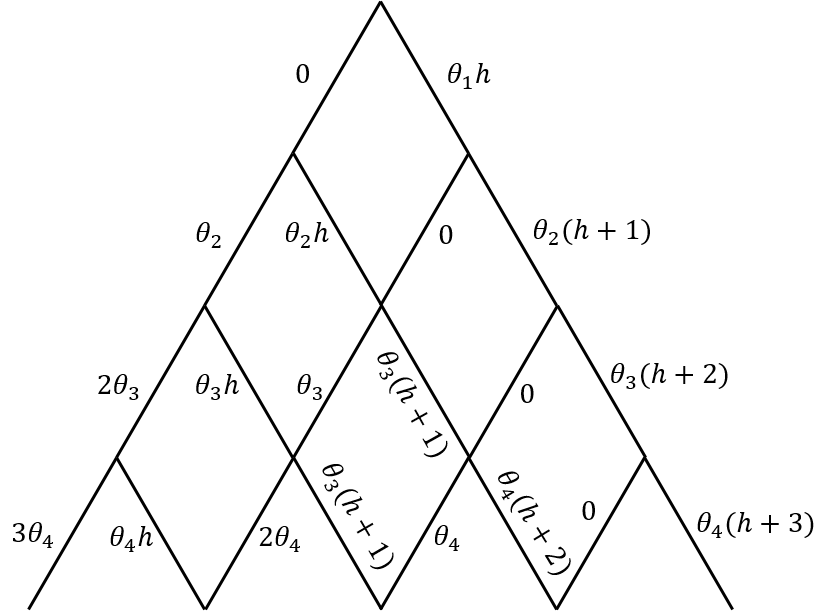}\includegraphics[width=4in]{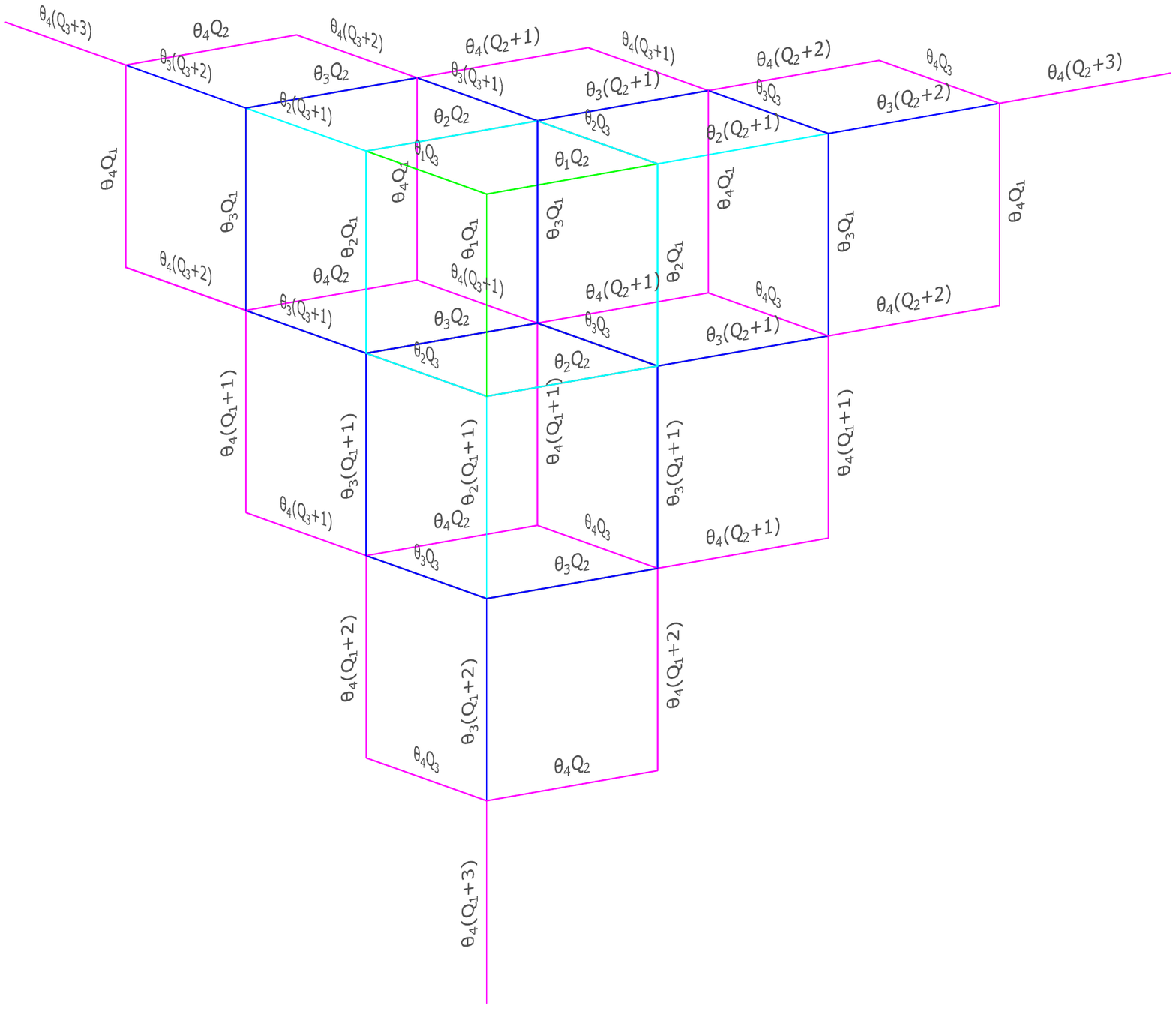}
\par\end{centering}
\caption{(left) 4-level Stackelberg with coalitions with 2 lanes (right) n-level
Stackelberg strategy graph for 3 lanes and 4 agents}
\end{figure}

\newpage
\begin{prop}
The computational complexity of the recursion for the vertical queue
case is $O(n(n+l)^{l})$: 

Counting the number of nodes at the final level is equivalent to finding
the number of $l$-combinations with repetitions:

$\left(\left(\begin{array}{c}
l\\
n
\end{array}\right)\right)=\left(\begin{array}{c}
l+n-1\\
n
\end{array}\right)=\frac{\prod_{i=1}^{l}(n+i)}{l-1}<\frac{(n+l)^{l}}{l-1}<(n+l)^{l}$

Since there are $n$ levels, the final complexity is $O(n(n+l)^{l})$.
\end{prop}
The recursion above needs to be solved for every partition $P\in\mathcal{P}$
to obtain all coalition values. The interaction between all the n-level
Stackelberg games with coalitions will be modeled as a cooperative
game. Cooperative games are complete-information games in which users
are allowed to form coalitions to improve their payoffs. In the absence
of coalitional externalities, cooperative games can be represented
in a Characteristic Function Form (CFF), which defines a Characteristic
Function Game (CFG). However, in the current problem there are externalities,
which translates into coalitional payoffs being dependent on which
other coalitions are formed. Thus, we will represent the game on a
partition function form (PFF) which defines a Partition Function Game
(PFG). \citep{ichiishi1981} shows that any strategic game as can
be represented as a partition function game. Thus, we will translate
the former strategic games as a single PFG.

Let $P=\{S_{1},\dots,S_{k}\}\in\mathcal{P}$ be a partition of $N$
such that $S_{i}\cap S_{j}=\emptyset\;\forall i\neq j$. We define
the partition function$v:\;2^{N}\times\mathcal{P}\rightarrow\Re$.
That is, $v(S,P)$ represents the value of coalition $S$ when the
partition $P$ is formed. $v(S,P)$ is in fact the sum of all the
valuations $v_{i}\;\forall i\in S,\forall S\in P,$ coming from the
optimal order resulting from the cooperation between agents belonging
to $S$ when the partition formed is $P$. The pair$<N,v>$defines
a partition function game. There is a coalition of special interest,
the grand coalition $S_{G}=N$, which is composed by all members of
the participant set.

A fundamental question in cooperative game theory is if this total
cooperation will eventually occur. This is desirable when the grand coalition
is the most efficient coalition. When the partition function game
arises from a strategic game, the grand coalition is always efficient
since the set of strategies of the grand coalition game includes all
the strategies available in the other subgames. In fact, the grand
coalition payoff is the shortest path on the graph defined by the
recursion (\ref{eq:recursion}) when $P=N$. Conversely, a coalition
can unanimously dissolve itself into singletons, since every member
can still choose the same set of strategies.

A fundamental characteristic of PFGs are the externalities that form
on a coalition by third coalitions merging or splitting. Literature
defines two types of externalities, positive and negative, which we
define next.
\begin{defn}
Positive (negative) externalities:
\[
\forall C,S,T\;|\;C\cap S\cap T=\emptyset,\;\forall\rho\in P(N-(S\cup T\cup C)):\;v(C;\{S\cup T,C\}\cup\rho)>(<)\;v(C;\{S,T,C\}\cup\rho)
\]
\end{defn}
Basically, a PFG displays positive externalities when two coalitions
$S$ and $T$ merging, increases the value of a third coalition $C$,
for any complementary partition $\rho$. Conversely, the externalities
are negative if the value of $C$ is decreased. It is easy to see
that this game has positive externalities. 

Let $\rho\in P(N\backslash(S\cup T\cup C))$. Suppose all agents forming
coalition $C$ precede those of coalitions $S$ and $T$. Then, the
merging of $S$ and $T$ has no influence on $v(C,\{C,S\cup T,\rho\})\;\forall\rho\in P(N\backslash(S\cup T\cup C))$
and the externality is zero. If agents forming $C$ go after $S$
and $T$, the externality can only be positive or zero since $S\cup T$
either causes some agents to join longer queues or keep the positions
if no improvement is possible, reducing the cost of the members of
$C$, which manage to advance some positions or none. However, if
members of $C$ are between those of $S,T$ or $S\cup T$, negative
externalities may occur. We provide an example next:
\begin{example}
The static problem with instance $N=\{13,2,14,41\},\;Q=\{4,1\}$ has
negative externalties. Let $S=\{1,4\}$, $T=\{3\}$,$C=\{2\}$. Then,
by solving the game with the recursion algorithm, we reach the case
of $0=v(C|\{S\cup T,C\})<v(C|S,T,C)=2$.
\end{example}

A useful property some PFG's exhibit is superadditivity, which is
defined next:

\begin{equation}
\forall S,T\subseteq N\;|\;S\cap T=\emptyset,\;\forall\rho\in N-(S\cup T)\;,v(S\cup T;\{S\cup T\}\cup\rho)\geq v(S;\{S,T\}\cup\rho)+v(T;\{S,T\}\cup\rho)\label{eq:def_SA}
\end{equation}

Superadditive means that if two coalitions $S$ and $T$ merge, their
total payoff is larger than when unmerged. For the sake of exposition,
the following counterexample shows that this game is not superadditive:
\begin{example}
$N=\{1,9,5,33\}$ and $h=3$. If $\rho=\{1,3\},\;S=\{2\},\;T=\{4\},$then
$v(S\cup T,\{S\cup T,\rho\})=15,\;v(S,\{S,T\}\cup\rho)=9,\;v(T,\{S,T\}\cup\rho)=33$.
\end{example}

PFG's stability concepts and characterizations generally focus on
games which have either positive or negative externalities \citep{Abe2016}
or exhibit superadditivity or convexity \citep{Hafalir2007}. This
is not the case of our problem. We found an exception in the literature,
which is the strong-core for PFG \citep{ChanderPFG2014}, which is defined
next:
\begin{defn}
Strong-core for PFGs \citep{ChanderPFG2014}

$(x_{1},\dots,x_{n})\in\Re^{n}\;|\;\forall P\in\mathcal{P},P=\{S_{1},\dots,S_{p}\}\neq[N],\;\exists S_{i}\in P,|S_{i}|>1\;|\;\sum_{j\in S_{i}}x_{j}\geq v(S_{i},P)\;and\;if\;P=[N],\;x_{i}\geq v(i;[N])\;\forall i\in N$
\end{defn}
The definition above states that for every partition which contains
non-singleton coalitions, exists at least one coalition of those coalitions
which is worse off than in the strong-core imputation $(x_{1},\dots,x_{n})$. Imputation is a  term used in game theory to denote the utility agents obtain from a coalitional agreement. Moreover, every agent in the all-singleton partition is worse off. Contrary to other core specifications found in the literature such as the $\alpha,\beta,\gamma,\delta$-cores
\citep{hartkurz1983}, the strong-core does not assume any coalition
structure for the complementary partition when a given coalition forms.
The solution concept is particularly useful for the treatment of the
status-quo partition, which, in our case, corresponds to the FCFS
queue allocation. If an imputation vector belongs to the strong-core,
then all the singleton coalitions belonging to the coalition made
of just singletons are better off by merging into the grand coalition.
Now the question that remains is to prove non-emptiness of the strong-core.
As shown above, our problem has both positive and negative externalities.
In order to the strong core to be non-empty, the PFG needs to satisfy
two conditions:

\begin{thm}
(\citep{ChanderPFG2014}, Corollary 5) A PFG with general externalities $<N,v>$ has a
non-empty strong-core if :
\end{thm}
\begin{enumerate}
\item $v$ is partially superadditive: $\forall P=\{S_{1},\dots,S_{m}\}\in\mathcal{P},\;|\;|S_{i}|>1\;\forall i=1,\dots,k\;|S_{j}|\;\forall j=k/+1,\dots,m\;k\leq m,\;\sum_{i}^{k}v(S_{i},P)\leq v(S,P^{\prime})\;P^{\prime}=P\backslash\{S_{1},\dots,S_{k}\}\cup\{\cup_{i=1}^{k}S_{i}\}$
\item and $<N,w^{\gamma}>$is balanced, where $w^{\gamma}(S)=v(S,\{S,[N\backslash S]\}),S\subset N$.
\end{enumerate}
Partial superadditivity is weaker than superadditive, which the game
does not satisfy. Partial superadditivity is trivially satisfied for
games with 3 or 4 players whenever the grand coalition is efficient.

\begin{conjecture}
The strong-core for the static queuing game as vertical queue is non-empty.
\end{conjecture}
We leave the result as a conjecture since it was not possible to prove
it. After having run a large number of simulations, we did not find
any counterexample, either. The values employed for the simulation
study are: $n\sim Unif(1,\bar{n}),\;\overline{n}\in[2,7],\;m\in[1,4],\;\theta\sim logn(\mu=2.16,\sigma=0.7),\;Q_{m}\sim Unif(1,4),\;Q_{i}\sim Unif(1,Q_{i+1})\;\forall i<m$.
The semi-empirical distribution used to draw individual Valuation
of Delay Savings is developed in \citep{Lloret-Batlle2016}.

Standard proof methods, such as direct proof used in operations research
games for Minimum Spanning Tree Games and Shortest Path games \citep{Borm2001},
did not prove successful. Neither did other approaches such as reduction
to market games. Moreover, the game did not prove to be convex, which
is a sufficient condition for non-emptiness of strong-core.

Instead, we are going to evaluate the inclusion of two imputations
which are generalizations of the Shapley value for partition function
games. These imputations satisfy the four properties of the original
Shapley value: efficiency, symmetry, additivity and null player. The
two imputations are presented next:
\begin{defn}
\citep{dohorde2007,Clippel2008} Externality-free value:
\begin{equation}
\phi_{i}^{free}(v)=\sum_{S\subseteq N}\zeta_{S}^{i}v(S,\{S\}\cup\{\{j\}\;|\;j\in N\backslash S\})\;\forall i\in N\label{eq:ext_free}
\end{equation}
Where:
\begin{equation}
\zeta_{S}^{i}=\begin{cases}
\begin{array}{cc}
\frac{(|S|-1)!(|N|-|S|)!}{|N|!} & if\;i\in S\\
-\frac{|S|!(|N|-|S|-1)!}{|N|!} & if\;i\notin S
\end{array}\end{cases}\label{eq:ext_free coeffs}
\end{equation}
\end{defn}
The $\zeta_{S}^{i}$ values arise from the reordering of the marginal
increment $v(S\cup\{i\})-v(S)\;\forall S\subseteq N\backslash\{i\}$
expression, often found in the Shapley value definition, in terms
of all the partitions $v(S)\;\forall S\subseteq N$.

This value represents that an agent leaving the grand coalition always
creates a new coalition, that is, a singleton. This value is inline
with the $\gamma$-core present in the strong core existence theorem and
we expect imputations from this value to generally belong to the strong-core.
The second imputation to test is:
\begin{defn}
\citep{McQuillin2009} McQuillin value:
\[
\phi_{i}^{McQ}(v)=\sum_{S\subseteq N}\zeta_{S}^{i}v(S,\{S,N\backslash S\})\;\forall i\in N
\]
This value entails that an agent always chooses an existing coalition.
\end{defn}
The following MILP will be used to test the feasibility of $\phi^{free}(v)$
and $\phi^{McQ}(v)$ in the strong-core. We add the following objective
function and modify the group rationality condition for the coalitions
which have non-singleton partitions $P\in\hat{\mathcal{P}}=\mathcal{P}\backslash\{[N]\cup\{N\}\}$.
We call this relaxation the$\epsilon$strong-core for PFG's, in line
with the $\epsilon$-core for characteristic function form games \citep{ShapleyShubik1966}.

\[
\min\epsilon
\]
\[
s.t.
\]

\begin{equation}
\sum_{i\in S_{j}}x_{i}\geq v(S_{j},P)-\epsilon-Mz_{jp}\;\forall S_{j}\in\ddot{P}\subseteq P,\forall P\in\hat{\mathcal{P}}\label{eq:score_group_rat_relax-1}
\end{equation}

\begin{equation}
x_{i}\geq v(\{i\},[N])-\epsilon\;\forall i\in N\label{eq:score_IR}
\end{equation}
\begin{equation}
\sum_{i\in N}x_{i}=v(N,\{N\})\label{eq:score_eff}
\end{equation}

\textbf{
\begin{equation}
\sum_{S_{j}\in\ddot{P}}z_{jp}\leq|\ddot{P}|-1\;\forall\ddot{P}\subseteq P\in\hat{\mathcal{P}}\label{eq:score_bin_bound-1}
\end{equation}
}
\begin{equation}
\epsilon\geq0,\;z_{jp}\in\{0,1\}\;\forall S_{j}\in\ddot{P}\subseteq P,\forall P\in\hat{\mathcal{P}}\label{eq:score_var_bounds}
\end{equation}
Where $\ddot{P}\cup\dot{P}=P\;|\;\ddot{P}\cap\dot{P}\neq\emptyset$
are the collections of non-singleton coalitions $\ddot{P}$ and singleton
coalitions $\dot{P}$ of every partition $P$. Essentially, the program
consists of the relaxed group rationality constraints (\ref{eq:score_group_rat_relax-1}),
the individually rational constraints (\ref{eq:score_IR}), the grand
coalition efficiency (\ref{eq:score_eff}). The $\epsilon$ variable
is the minimal slack for the most constraint coalition $S_{j}\in\ddot{P}\subseteq P$
necessary to make the problem feasible. The binary terms $z_{jp}$
present in (\ref{eq:score_group_rat_relax-1}) and (\ref{eq:score_bin_bound-1})
enforce that at least one non-singleton coalition $S_{j}\in\ddot{P}\subseteq P$
for every $P\in\hat{\mathcal{P}}$ to be group rational.

The settings of this simulation are the same ones than for strong-core
non-emptiness evaluation. For each of these settings, 250 experiments
are run. The next tables show the percent of instances where the imputation
was found in the strong-core:

\begin{table}[h]
\begin{minipage}[t]{0.45\columnwidth}%
\begin{center}
\begin{tabular}{|c|c|c|c|}
\hline 
$\phi^{free}:\;\overline{n}\backslash M$ & 2 & 3 & 4\tabularnewline
\hline 
2 & 100.0 & 100.0 & 100.0\tabularnewline
\hline 
3 & 100.0 & 97.2 & 98.0\tabularnewline
\hline 
4 & 96.0 & 85.2 & 88.4\tabularnewline
\hline 
5 & 87.6 & 75.6 & 72.4\tabularnewline
\hline 
6 & 84.0 & 67.6 & 64.0\tabularnewline
\hline 
7 & 74.4 & 52.8 & 61.6\tabularnewline
\hline 
\end{tabular}
\par\end{center}%
\end{minipage}\hfill{}%
\begin{minipage}[t]{0.45\columnwidth}%
\begin{center}
\begin{tabular}{|c|c|c|c|}
\hline 
$\phi^{McQ}:\;\overline{n}\backslash M$ & 2 & 3 & 4\tabularnewline
\hline 
2 & 100.0 & 100.0 & 100.0\tabularnewline
\hline 
3 & 100.0 & 97.2 & 98.0\tabularnewline
\hline 
4 & 96.0 & 85.2 & 88.4\tabularnewline
\hline 
5 & 88.0 & 75.6 & 72.4\tabularnewline
\hline 
6 & 83.6 & 67.6 & 64.0\tabularnewline
\hline 
7 & 74.4 & 53.2 & 62.0\tabularnewline
\hline 
\end{tabular}
\par\end{center}%
\end{minipage}

\caption{Percent of experiments whose imputations are in the strong-core.}
\end{table}

We observe that both imputations provide very similar percentages
of inclusion into the strong-core, the only differences being due
to numerical errors from the solution algorithm. As expected, the
larger the vehicle set and the larger the number of lanes, the more
the instances of not belonging to the strong core. This result seems
to be weaker when increasing the number of lanes than when increasing
the maximum platoon size. Both imputations having identical percents
suggest that a certain degree of symmetry exists in the violation
of the strong-core in the coalition formation process, for both all-singleton
blocking coalitions and all-maximum-cardinality complementary blocking
coalitions. This can provide some insights for the non-emptiness proof
of this vertical queue model.

\section{Parallel horizontal queues: Dynamic problem}

Queues in transportation systems have a dynamic nature: they continuously
receive arrivals and dispatch departures. Moreover, these queues are
``horizontal'', that is, queue length has a physical dimension.
Modeling such dynamic process as a succession of static queuing
problems may lead to strong inter-temporal inefficiencies, coalitional
stability and violation of individual rationality. One of the solutions,
employed in \citep{lloretbatlle2017ISTTT} is to set up a reserve
price which prevents low value vehicles from having too much present
savings in detriment of further high value vehicles. Another alternative
would be to add a terminal cost at the leaves of the static problem
tree, however this would require a transfer of payment from former
agents to new agents as well as knowledge of further arrivals and
their values, which would difficult the dynamic budget balancedness
and the efficiency of the system. In this particular problem, we will
stick to a control solution which is always budget balanced and efficient,
aiming an eventual V2V decentralized implementation.

We model again a link with $M$ lanes, but this time with length $\lambda$. Vehicles
enter the link upstream at a constant speed $v_{a}$. There is a bottleneck
downstream with a constant outflow $q_{out}$ which corresponds to
a headway $h_{q}$, spacing $s_{q}$ and speed $v_{q}<v_{a}$. Every
time an event happens, vehicle communicate to each other their value
of time and positions, including the vehicles at the back of the queue.
An imputation which satisfies the minimal $\epsilon$-strong core
is found and vehicles are assigned their new lanes. To minimize excessive
perturbations due to the lane changes, only vehicles whose incorporation
to the back of the queue is imminent will execute the lane change.
The rest of vehicles will only execute the lane change once their
incorporation becomes imminent. Naturally, the target lane can later
change if there are further stability optimizations being executed
due to new events happening.

Vehicles participate in lane-changing optimizations only as long as
their incorporation to the back of the queue is not imminent. Once
a vehicle is queued, they do not participate in other cooperative
lane changes and are supposed to stay in the queue. The exchange optimization
is run at every instant when there is a significant event. We define
an event as the arrival of new vehicle to the link or imminent proximity
of a moving vehicle to the back of the queue, or any external unpredictable
event which could be detected by any of the vehicles. Events happen
at time instants $t$, called epochs.

With this in mind, and using Newell's simplified car-following model,
vehicle's $i$ predicted delay at epoch $t$ is the maximum quantity
of two situations: arriving to the bottleneck downstream undelayed
at free flow speed or being queued behind its predecessor:

\begin{equation}
d_{i}^{t}(S,P)=\max\{t_{i-1,t}^{dep}(S,P)+\frac{s_{q}}{v_{q}},a_{i}\}-a_{i}
\end{equation}

Since vehicles may participate in multiple optimizations, the utility
specification is composed of the predicted total cost at epoch $t$
since its arrival to the system minus the price charged at the optimization
executed during epoch $t$ minus the accumulated price charged to vehicle
$i$ until $t$. $i$'s imputation from being in coalition $S$ and
partition $P$ is:

\begin{equation}
x_{i}^{t}(S,P)=v_{i}^{t}(S,P)-p_{i}^{t}(S,P)-\pi_{i}^{t-1}\;\forall i\in I_{t},\forall S\subseteq P,\forall P\in\hat{\mathcal{P}}
\end{equation}

\begin{equation}
v_{i}^{t}(S,P)=-c_{i}^{t}(S,P)=-\theta_{i}d_{i}^{t}(S,P)
\end{equation}

With the imputations and valuations being defined, the dynamic $\epsilon$-strong core program is defined by replacing $\forall i \in N^t$, $x_i^t$ and $v_i^t$ into equations (\ref{eq:score_IR} - \ref{eq:score_bin_bound-1}). Naturally, the $\epsilon$ term in equations (\ref{eq:score_var_bounds}) and in the objective function will be replaced at each program $t$ by the corresponding $\epsilon^t$.

%\[
%\min\epsilon^{t}
%\]
%\[
%s.t.
%\]
%
%\begin{equation}
%\sum_{i\in S_{j}}x_{i}^{t}\geq v(S_{j},P)-\epsilon-Mz_{jp}\;\forall S_{j}\in\ddot{P}\subseteq P,\forall P\in\hat{\mathcal{P}}\label{eq:score_group_rat_relax-1-1}
%\end{equation}
%
%\begin{equation}
%x_{i}^{t}\geq v(\{i\},[N^{t}])-\epsilon_{t}\;\forall i\in N^{t}\label{eq:score_IR-1}
%\end{equation}
%\begin{equation}
%\sum_{i\in N}x_{i}^{t}=v(N^{t},\{N^{t}\})\label{eq:score_eff-1}
%\end{equation}
%
%\textbf{
%\begin{equation}
%\sum_{S_{j}\in\ddot{P}}z_{jp}\leq|\ddot{P}|-1\;\forall\ddot{P}\subseteq P\in\hat{\mathcal{P}}\label{eq:score_bin_bound-1-1}
%\end{equation}
%}
%\begin{equation}
%\epsilon^{t}\geq0,\;y_{jp}\in\{0,1\}\;\forall S_{j}\in\ddot{P}\subseteq P,\forall P\in\hat{\mathcal{P}}\label{eq:score_var_bounds-1}
%\end{equation}
%
%The program is identical to the static case one, it is included for
%the sake of completion.
%

Parallel horizontal queues cannot benefit from the polynomial structure
employed in the previous section. Vehicles' costs do not depend this
time only on the queue length in front of them, but on the dispatching
times of the vehicles' downstream. Since the departure times depend
on the actual sequence of queued vehicles, a particular state is now
defined by the sequence of vehicles in front of them and therefore
the whole queuing tree needs to be explored. This increases the computational
complexity of the problem. For this reason, we will limit the number
of participant agents of every optimization to six. Any additional
vehicles upstream will stay outside of the exchange and continue advancing
through the link in a FCFS basis. The model with the full exponential tree is general enough to include jockeying strategies. That is, agents are allowed to switch queues after having joined a queue. We rule out this possibility in this paper for simplicity in exposition.

\begin{prop}
The computational complexity of the recursion for the horizontal queue
case is $O(l^{n})$:

For the horizontal queue case, all the tree histories need to be explored.
This defines a $l$-ary tree with $n$ levels. The last level has
$O(l^{n})$ nodes.
\end{prop}

The exchange of information in positions also serves to define which
lane changes are possible and which ones are obstructed. In our current
formulation, if some lane change is not possible at a particular epoch
$t$, the cost of it equivalent branch is set to $\infty$ and that
strategy does not get explored. However, this is not implemented here.

We explore next the core stability of the dynamic problem as horizontal
queues. We run 6 1-hour long simulations for each of the scenarios defined by: $\lambda=200\;m,\;L\in\{2,3\},\;q_{in}\in\{360,540,720\}\;veh/h/lane,$ $\;q_{out}=900\;veh/h/lane,\;\theta\sim logn(\mu=2.16,\sigma=0.7)$.
Arrivals arise from a binomial distribution for implementation easiness,
but the simulation is still event-based: both time and distance are
continuous. The simulation is coded in MATLAB and the MILP programs
are solved with Gurobi 6.0.5.

The next table (left) shows the percent of optimizations which are
contained in the strong-score. We observe that for equal vehicle inflow,
increasing the number of lanes increases core stability. This can
be explained by the incoming platoon gets split into more lanes and
their queues and interactions are smaller. Furthermore, increasing
the incoming flow seems to increase instability, mostly due to a natural
increase of complexity in the strategic interaction between agents.
The table on the right displays the average ratio between $\epsilon$
and the average vehicle cost per optimization.
This ratio is useful to compare the magnitude of the slack term, i.e.
amount of utility that has to be transferred to a blocking coalition,
with regard to the average cost of the participating agent. In line
with the previous analysis, for equal inflow, a larger number of lanes
decreases the magnitude of the instability. However, increasing the
inflow seems to decrease the ratio, probably due to a larger increase
in average vehicle cost.

\begin{table}[h]
\begin{minipage}[t]{0.45\columnwidth}%
\begin{center}
\begin{tabular}{|c|c|c|}
\hline 
$\;q_{in} (veh/h) \backslash M$ & 2 & 3\tabularnewline
\hline 
360 & 91.5 & 99.6\tabularnewline
\hline 
540 & 88.5 & 95.1\tabularnewline
\hline 
720 & 78.7 & 93.2\tabularnewline
\hline 
\end{tabular}
\par\end{center}%
\end{minipage}\hfill{}%
\begin{minipage}[t]{0.45\columnwidth}%
\begin{center}
\begin{tabular}{|c|c|c|}
\hline
$q_{in} (veh/h) \backslash M$ & 2 & 3\tabularnewline
\hline 
360 & 0.29 & 0.17\tabularnewline
\hline 
540 & 0.11 & 0.16\tabularnewline
\hline 
720 & 0.04 & 0.02\tabularnewline
\hline 
\end{tabular}
\par\end{center}%
\end{minipage}

\caption{(left) \% of strong-core stable optimizations. (right) ratio $\epsilon/av.cost$ }
\end{table}

\section{Conclusions and further research}

This article presented a new collaborative control mechanism for freeways
and parallel queue facilities. Under this control scheme, agents observe
predicted future delays per lane (or queue) and are allowed to collaborate
to change lanes such that the total travel cost of their platoon is
minimized. High VOT vehicles can pay low VOT vehicles to switch to
a more congested lane while they can stay in the same lane or switch
to another lane with less vehicles in front. The underlying cooperative
principle is the strong-core for partition function games. While aimed
as a decentralized, distributed control, the present paper assumes
a centralized optimization to evaluate the economic efficiency and
stability without excessive technicalities.

The control policy has been first explored as a simpler vertical queue
model. In this case, the strategic interaction between users structure
forms a tree-like structure which is of polynomial-time complexity.
While not proved, simulation results suggest that the problem may
be strong-core stable. In addition, we have tested two generalizations
of the Shapley value for partition function games which we found to
generally be strong-core stable.

In the next section, we modeled the control policy as a dynamic horizontal
queue. We have observed that the policy is generally strong-core stable
except for situations when there are sharp increases in the incoming
platoon size. However, it is computationally intensive. Further distributed
optimization techniques should be used to make it applicable for an
eventual real-world implementation. Alternatively, designing an approximation
algorithm for the strong-core optimization would serve.

As further research we point out the following lines. We believe that
developing a formal non-emptiness proof for the vertical case would
represent a strong result in cooperative game theory. Also, from a
theoretical point of view, exploring the dynamic vertical queue case
would be interesting in the sense that dynamic applications in partition
function games have never been explored outside of coalition formation
in static settings. Concerning the dynamic horizontal queue, we will
further explore unstable instances to better determine the source
of instability and better understand the problem. Eventually, modeling
such policy with a commercial microsimulation software, would provide
insights on the efficiency increases of a real world situation, as
well as including more realism and efficiency losses due to lane changing
obstruction.

\bibliographystyle{plain}
\bibliography{library}

\end{document}